\documentclass[a4paper,11pt]{article}
\usepackage{pos}
\usepackage{float}
\usepackage{subcaption}
\usepackage{multirow}

\title{\begin{flushright}
\small{WUB/25-01\\}
\end{flushright}
Static-light meson spectroscopy with optimal distillation profiles}
\ShortTitle{Static-light meson spectroscopy with optimal distillation profiles}

\author*[a]{Laura Struckmeier}
\author[a]{Roman Höllwieser}
\author[a]{Francesco Knechtli}
\author[a]{Tomasz Korzec} 
\author[b]{Michael Peardon}
\author[a]{Juan Andrés Urrea-Niño}

\affiliation[a]{Department of Physics, University of Wuppertal, Gaußstrasse 20, 42119 Germany}

\affiliation[b]{School of Mathematics, Trinity College Dublin, Ireland}
	
\emailAdd{lstruckmeier@uni-wuppertal.de}
\emailAdd{hoellwieser@uni-wuppertal.de}
\emailAdd{knechtli@uni-wuppertal.de}
\emailAdd{korzec@uni-wuppertal.de}
\emailAdd{mjp@maths.tcd.ie}
\emailAdd{urreanino@uni-wuppertal.de}

\abstract{The spectrum of static-light and static-charm mesons is studied using optimized distillation in two different $N_{\rm{f}} = 3 + 1$ QCD ensembles with pion masses of $m_{\pi} \approx 800 \, \text{MeV}$ and $m_{\pi} \approx 420 \,\text{MeV}$ and a heavy (charm) quark. Local and derivative-based operators are used to access states of multiple quantum numbers. The use of optimal profiles is shown to improve the overlap with the energy states compared to standard distillation.}

\FullConference{The 41st International Symposium on Lattice Field Theory (LATTICE2024)\\
 28 July - 3 August 2024\\
Liverpool, UK\\}

\begin{document}
\maketitle

\section{Introduction}
Investigating B-mesons gives access to several Standard Model parameters. Precise lattice calculations require good creation operators for the heavy-light mesons. A main issue of these B-meson calculations in lattice QCD is given by the heavy quark mass, which introduces a different scale than the one of the light quark and thus it makes the system computationally expensive. One way to handle this problem is using \textit{Heavy Quark Effective Theory} (HQET) \cite{sommer2010introduction,michele2005alpha}, where the heavy quark is considered to be infinitely heavy so that the static-light meson is given as the static limit of B-mesons. Since the static-light meson represents the leading term in HQET, investigating B-mesons serves as a test for HQET. Another application of static-light mesons is (hybrid) string-breaking, since two static-light mesons can be a product of string-breaking. It is therefore necessary to know the static-light meson spectrum, especially with angular momentum, for the hybrid string-breaking. Furthermore, studying the $B^\ast \pi$ excited state contamination in the B-meson spectrum can give deeper insight into overlaps and Heavy Meson Chiral Perturbation Theory \cite{bar2023bpiexcitedstatecontaminationlattice}. In this work, the improved distillation technique is applied to the system of one static-light and also static-charm meson, studying the performance of the optimal distillation profiles compared to standard distillation to obtain a precise estimation of the static-light (/-charm) spectrum of radial and orbital excitations. Results in two different $N_{\rm{f}} = 3 + 1$ QCD ensembles of $m_{\pi} \approx 800 \, \text{MeV}$ and $m_{\pi} \approx 420 \,\text{MeV}$ and an almost physical charm quark are shown to study the dependence on the pion mass.

\section{Construction of the operators}
Due to heavy quark spin symmetry in the static limit, the static quark can be treated solely as a color source without spin \cite{Michael_1998,Foley_2007}. Thus, the states are labeled according to the light-quark spin. The relevant lattice symmetry group is the double cover of the full cubic group $O_h^D$. To construct operators that transform according to the fermionic irreducible representations (irreps) of this group \cite{altmann1994point}, one can start by building a representation from a bispinor which is reducible into $G_1^+ \oplus G_1^-$ \cite{Peskin:1995ev}. The operators are then given by the projection onto these irreps. The best way to access further fermionic irreps is to introduce a lattice spatial covariant derivative, which transforms like $T_1^-$. Table \ref{tab:SubductionTable} shows examples of the resulting operators in the Euclidean chiral basis and their corresponding continuum angular momentum $j$ with parity $P$, subduced into the irreps of $O_h^D$ that are considered in this work. These operators can be related to the ones used in \cite{Michael_1998} or \cite{Jansen_2008} for example, by taking proper linear combinations. The spatial gauge links used here are 3D APE-smeared, while the temporal gauge links are HYP2-smeared, as described in more detail in \cite{Höllwieser_2020}.

\begin{table}[H]
        \centering
        \begin{tabular}{|c|c|c|c|} \hline
            $j^P$ & Irrep & Spectral notation & Example operators \\\hline\hline
            $1/2^+$ & $G_1^+$ & $S$ & $\left(\psi_1-\psi_3 \right)$, $\left[\nabla_1 (\psi_1 + \psi_3 ) + i\nabla_2 (\psi_1 + \psi_3) - \nabla_3 (\psi_2 + \psi_4)\right]$ \\\hline
            $1/2^-$ & $G_1^-$ & $P_{1/2}$ & $\left(\psi_1+\psi_3 \right)$, $\left[\nabla_1 (\psi_1 - \psi_3 ) + i\nabla_2 (\psi_1 - \psi_3) - \nabla_3 (\psi_2 - \psi_4)\right]$ \\\hline
            $3/2^-$ & $H^-$ & $P_{3/2}$ & $\left[2\nabla_1 (\psi_1-\psi_3) - i\nabla_2 (\psi_1-\psi_3) + \nabla_3 (\psi_2-\psi_4)\right]$ \\\hline
        \end{tabular}
        \caption{Relation of the irreps of $O_h^D$ to the continuum angular momenta $j$ with parity $P$, corresponding spectral notation and example operators, that are used here.}
        \label{tab:SubductionTable}
    \end{table}

\section{Static-light mesons with improved distillation}
To reduce the excited-state contamination in the correlation functions, the light quark is smeared using the improved distillation technique \cite{JuanPaper,Höllwieser_2023}. The idea of distillation \cite{Peardon_2009} is to project the quark fields onto the smaller eigenspace of the 3D gauge covariant lattice Laplacian, which then enables the exact evaluation of the smeared light quark propagator. Using a matrix $V(t)$ whose columns contain the $N_v$ lowest eigenmodes of the 3D gauge covariant Laplacian at time $t$ such that $\nabla^2(t)v_i(t)=\lambda_i(t)v_i(t)$, the quark propagators are replaced by the (static) perambulators $\tau$ and $\tau^{\text{(stat)}}$:
\begin{align*}
           D^{-1}_{\alpha\beta}(t_1,t_2) \quad &\longrightarrow \quad  \tau_{\alpha\beta}(t_1,t_2) = V^{\dagger}(t_1)D^{-1}_{\alpha\beta}(t_1,t_2)V(t_2)
           \\ \mathcal{P}(\Vec{x};t_1,t_2) \quad &\longrightarrow \quad \tau^{\text{stat}}(\Vec{x};t_1,t_2) = V^{\dagger}(\Vec{x};t_1)\mathcal{P}(\Vec{x};t_1,t_2) V(\Vec{x};t_2) \, ,
\end{align*}
where $\mathcal{P}(\Vec{x};t_1,t_2)$ is the temporal Wilson line that represents the static quark propagation in time and $\alpha, \beta$ denote the spin components. To improve on that method, one can introduce a profile function $\rho_i(t)\equiv\rho(\lambda_i(t))$ that modulates the contribution of each $v_i$. In the case of the local static-light meson operator, the correlation function is given by
\begin{align}\label{CorrelationLocal}
    &C(\Vec{x};t_2,t_1) = -\Big\langle \sum\limits_{i,j} \rho_i(t_2)\rho_j(t_1) \, \tau_{ji}^{\text{stat}}(\Vec{x};t_1,t_2) \, \left(\text{Tr}_{\text{spin}}[\Gamma\tau_{ij}(t_2,t_1)]\right) \Big\rangle_{\text{gauge}} \, ,
\end{align}
where $\Gamma$ is a $4\times 4$ matrix that picks the needed spin components of the light perambulator $\tau$. By choosing $N$ different profiles, in this work $N=7$ Gaussian profiles, a GEVP \cite{LuW,Sommer} can be solved to extract the energy eigenstates. Note that the correlation function of the derivative-based operators is given by a sum of terms like in \eqref{CorrelationLocal} with $\tau^{\text{stat}}$ replaced by $(\nabla_k V)^{\dagger}\mathcal{P} \nabla_k V$, where $\nabla_k$ is a gauge covariant derivative in  spatial direction $k$. \\

In the following, results of static-light and static-charm meson measurements of operators projected onto the fermionic irreps in two different $N_{\rm{f}}=3+1$ QCD ensembles are presented. They differ by their pion mass and have been generated using the action of \cite{action}. A1 was generated at the physical $SU(3)$ flavor symmetric point. Further details can be found in table \ref{tab:EnsembleParams} and in \cite{Höllwieser_2020}. For the statistical analysis, the \textit{pyerrors} library \cite{JOSWIG2023108750} is used.
\begin{table}[H]
       	\centering
        \begin{tabular}{c||c|c|c|c|c|c}
             & $L^3\times T$ & $a \, [\text{fm}]$ & $m_{\pi} \,[\text{MeV}]$ & $N_v^{\text{(light)}}$ & $N_v^{\text{(charm)}}$ & $N_{\text{{confg}}}$ \\\hline\hline
            A1 & $32^3 \times 96$ & $0.05359$ & $420$ & $100$ & $200$ & $4000$ \\
            A1h & $32^3 \times 96$ & $0.0690$ & $800$ & $200$ & $200$ & $2000$ \\
        \end{tabular}
        \caption{Ensemble parameters.}
        \label{tab:EnsembleParams}
\end{table}

\begin{figure}
\centering
\begin{subfigure}{0.49\textwidth}
            \includegraphics[width=\textwidth]{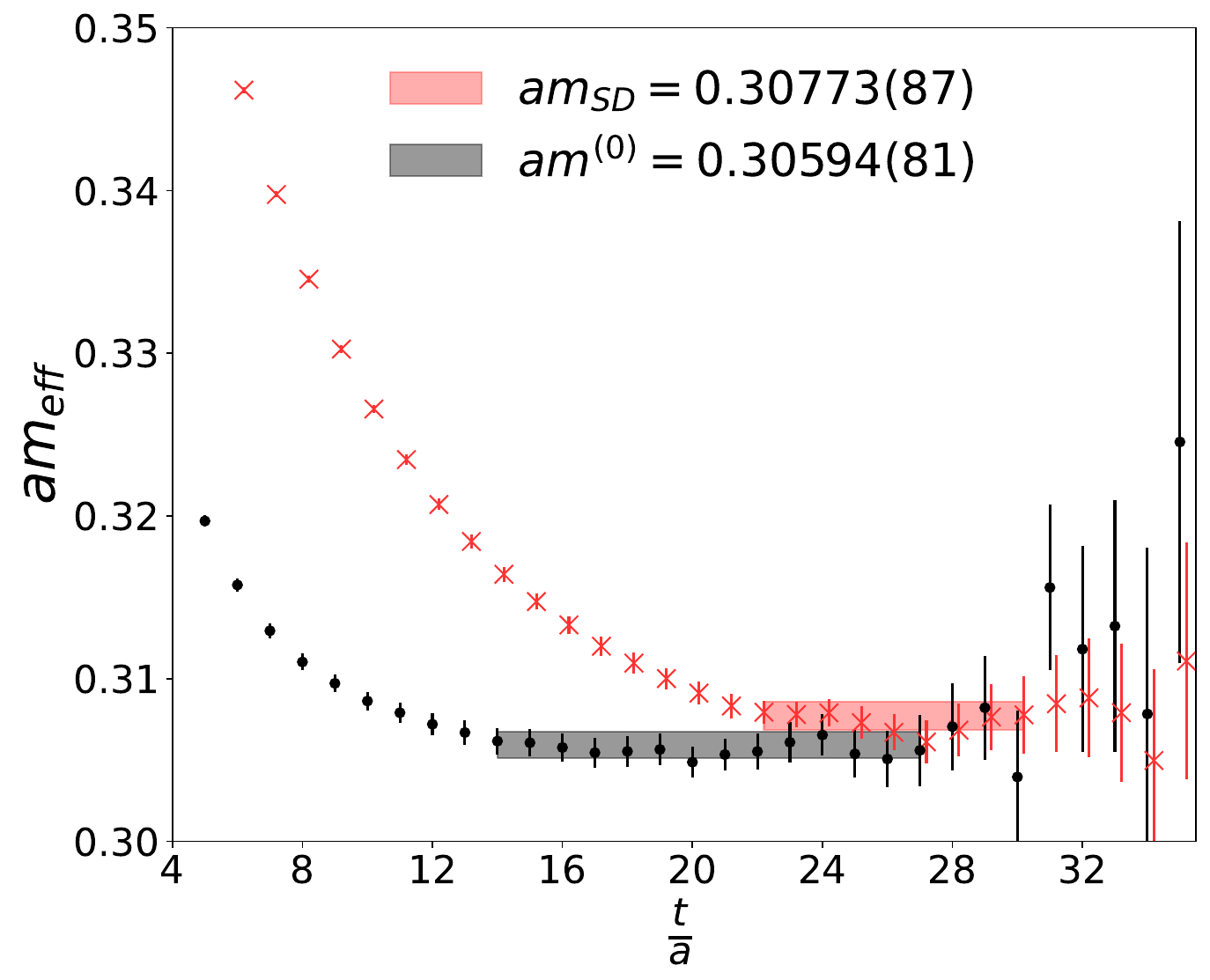}
    \caption{The standard distillation curve was obtained using $N_v=100$ eigenvectors.}
    \label{fig:first}
\end{subfigure}
\hfill
\begin{subfigure}{0.49\textwidth}
            \includegraphics[width=\textwidth]{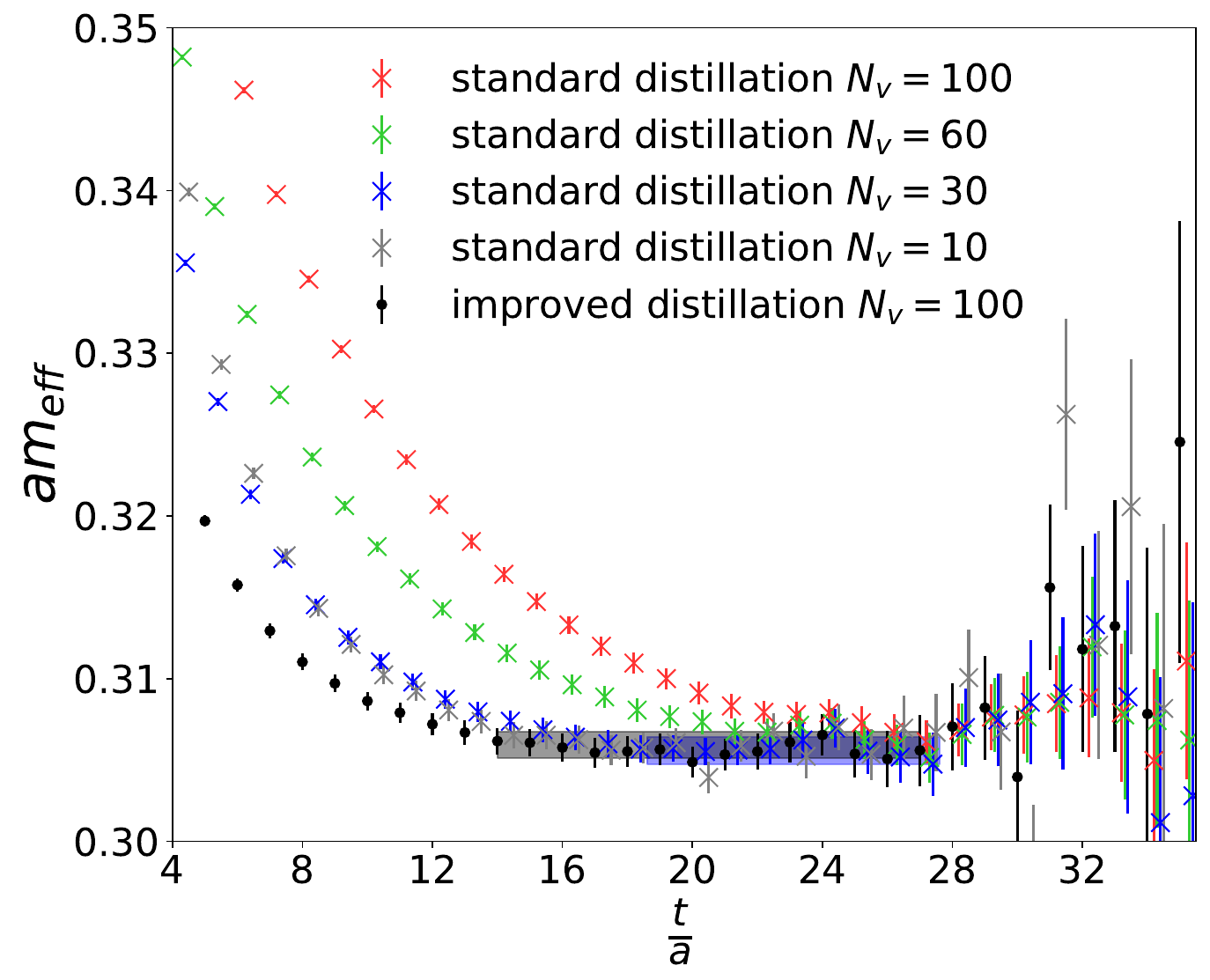}
    \caption{Different numbers of eigenvectors for the standard distillation.}
    \label{fig:second}
\end{subfigure}
\caption{Comparison of the effective ground state mass between standard distillation (colored x's) and improved distillation (black dots) obtained using $N_v=100$ eigenvectors for the ground state of the static-light meson on the A1 ensemble. The mass plateaus and their errors (light shaded bands) are obtained by a correlated linear fit to the logarithm of the effective mass.}
\label{fig:Comparison}
\end{figure}

\begin{figure}
\centering
\begin{subfigure}{0.49\textwidth}
            \includegraphics[width=\textwidth]{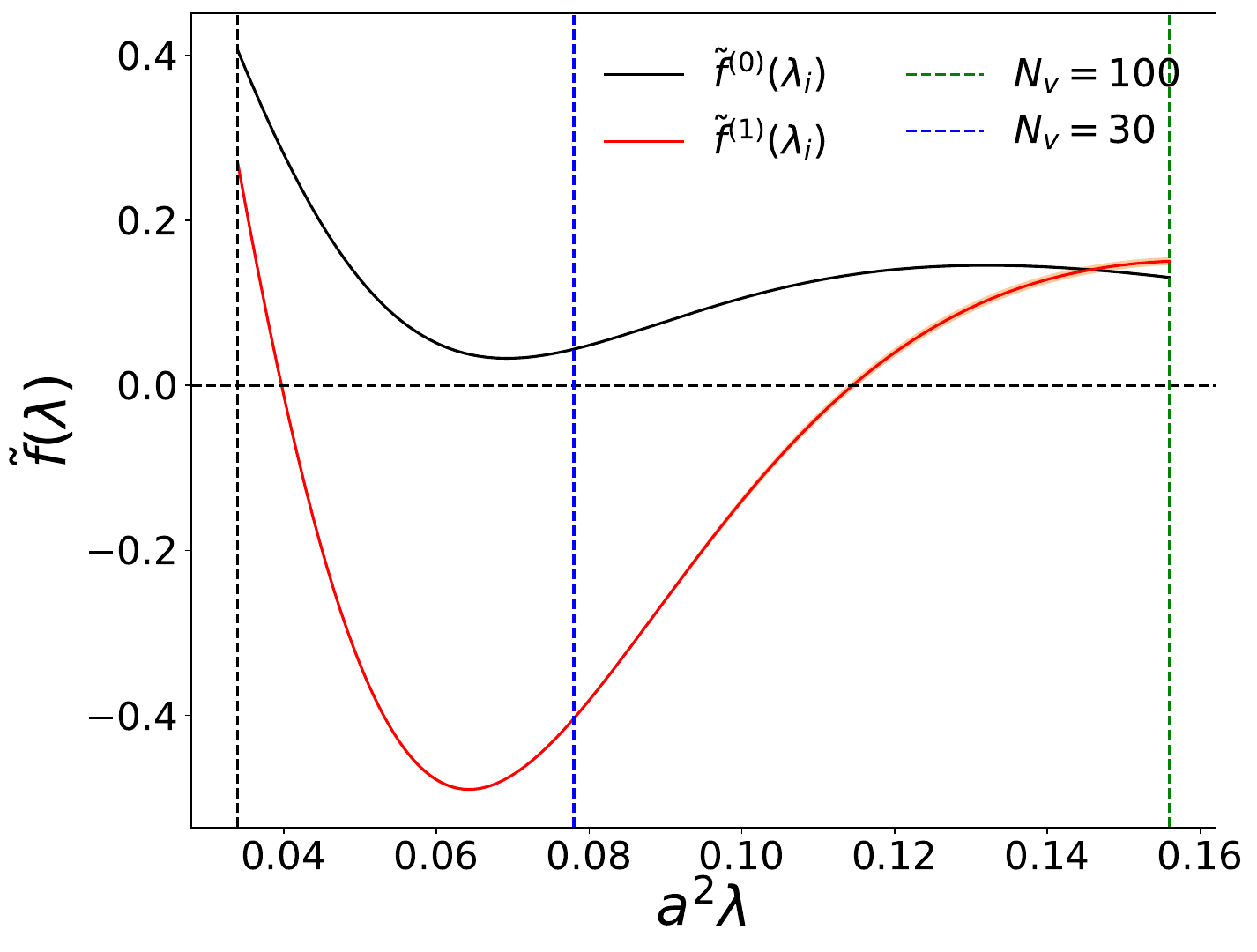}
    \caption{A1 ensemble.}
    \label{fig:first}
\end{subfigure}
\hfill
\begin{subfigure}{0.49\textwidth}
            \includegraphics[width=\textwidth]{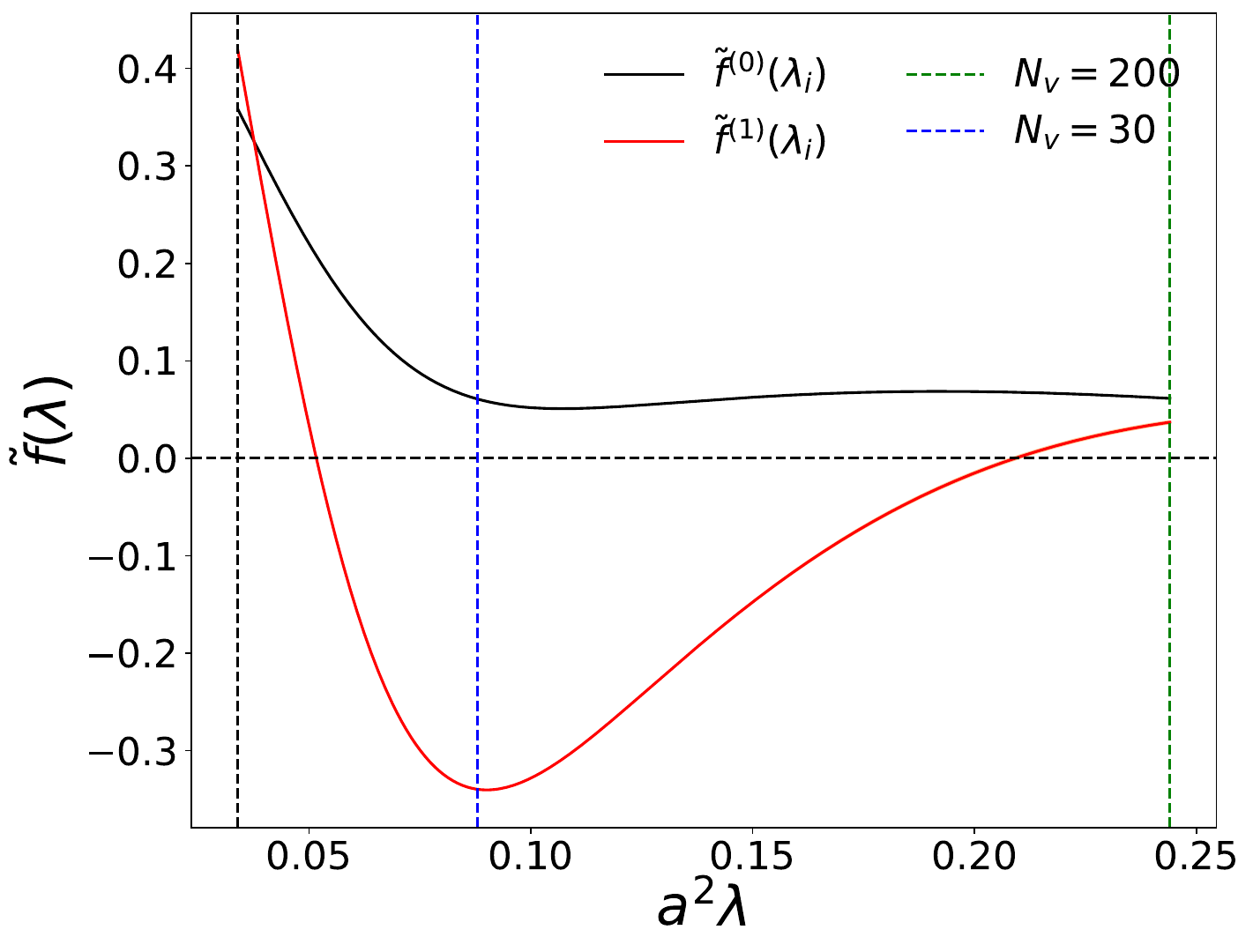}
    \caption{A1h ensemble.}
    \label{fig:second}
\end{subfigure}
\caption{Optimal meson distillation profiles $\Tilde{f}(\lambda)$ for the ground state (black) and first excited state (red) of the static-light meson.}
\label{fig:Profiles}
\end{figure}

Figure \ref{fig:Comparison} shows the ground state effective mass of the static-light meson on the A1 ensemble, depending on the time in lattice units, obtained by applying improved distillation (black dots) and standard distillation (colored x's). Including the optimal meson distillation profiles leads to a high suppression of excited state contamination, when using the same number of Laplacian eigenvectors (here $N_v=100$) for both standard distillation and improved distillation, as can be seen in the left panel of the figure. In the right panel, further effective masses are included that were obtained using standard distillation with fewer eigenvectors, which improves the mass estimation up to $N_v=30$. Comparing the improved distillation result (again with $N_v=100$) with the cleanest standard distillation effective mass (blue ($N_v=30$) or grey ($N_v=10$)), it can be seen that the improved distillation still shows less excited state contamination, resulting in a broader plateau and thus a smaller systematical error. Both plots confirm that including the optimal meson distillation profiles reduces the excited state contamination in the mass estimation for the static-light meson. \\
The optimal meson distillation profiles give further insights into the structure of the states. Figure \ref{fig:Profiles} shows examples for the ground and first excited state of the static-light meson depending on the Laplacian eigenvalues in lattice units for the A1 ensemble (left) and the A1h ensemble (right). The profiles show more structure the higher the state. Furthermore, larger eigenvalues still have a non-negligible contribution and are needed to resolve excited states. These results confirm the gain in using improved distillation for the static-light systems. According to these results, it is also used in the subsequent analysis to enhance the mass plateaus.

\section{Static-light and static-charm meson spectrum}
\begin{figure}[H]
\centering
\begin{subfigure}{0.49\textwidth}
            \includegraphics[width=\textwidth]{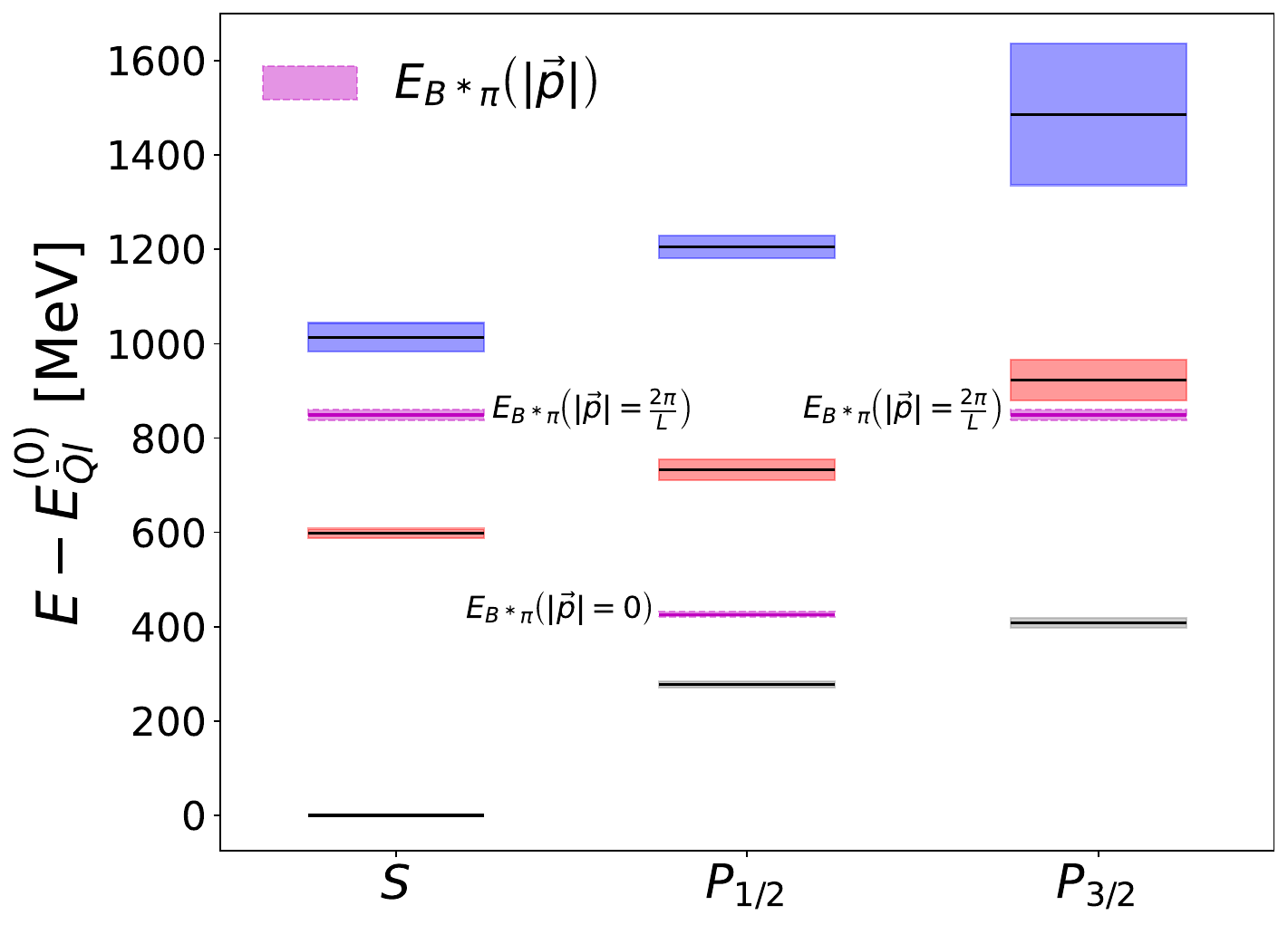}
    \caption{A1 ensemble.}
    \label{fig:first}
\end{subfigure}
\hfill
\begin{subfigure}{0.49\textwidth}
            \includegraphics[width=\textwidth]{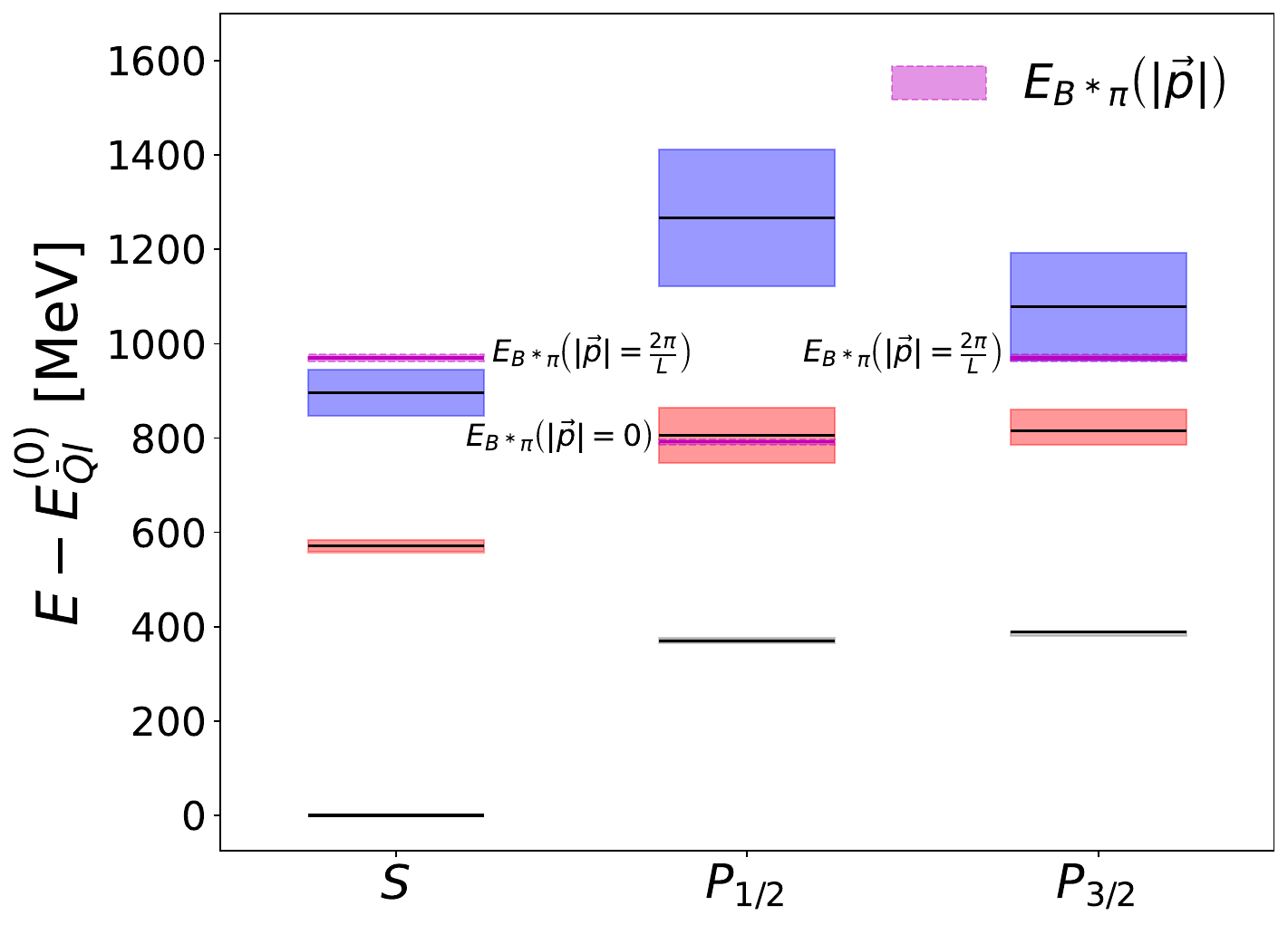}
    \caption{A1h ensemble.}
    \label{fig:second}
\end{subfigure}
\caption{Static-light meson spectrum with accessible radial and orbital excitations, obtained using improved distillation and non-interacting $B^{\ast}\pi$-states (magenta).}
\label{fig:StatLightSpectrum}
\end{figure}

\begin{figure}[H]
\centering
\begin{subfigure}{0.49\textwidth}
            \includegraphics[width=\textwidth]{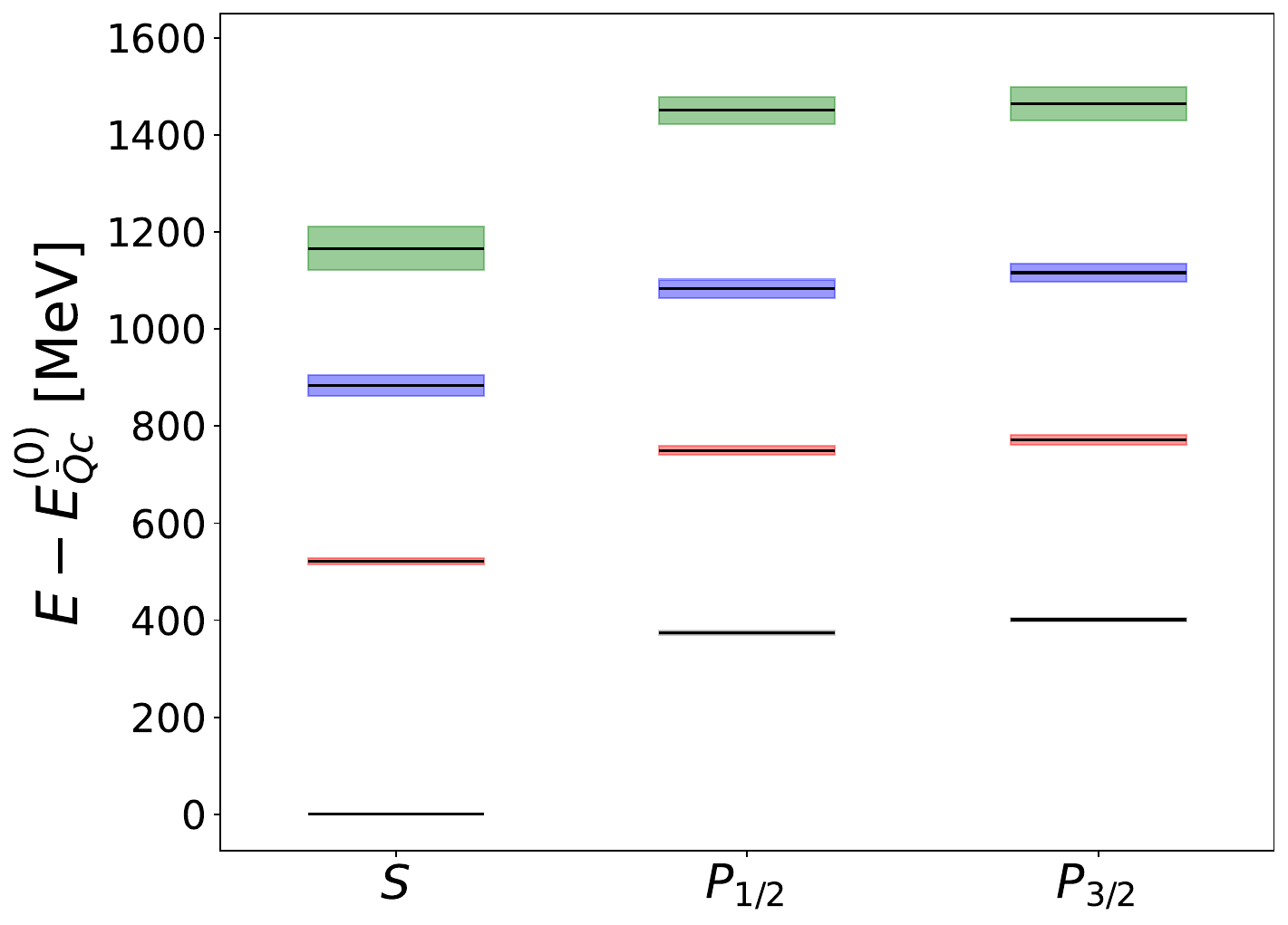}
    \caption{A1 ensemble.}
    \label{fig:first}
\end{subfigure}
\hfill
\begin{subfigure}{0.49\textwidth}
            \includegraphics[width=\textwidth]{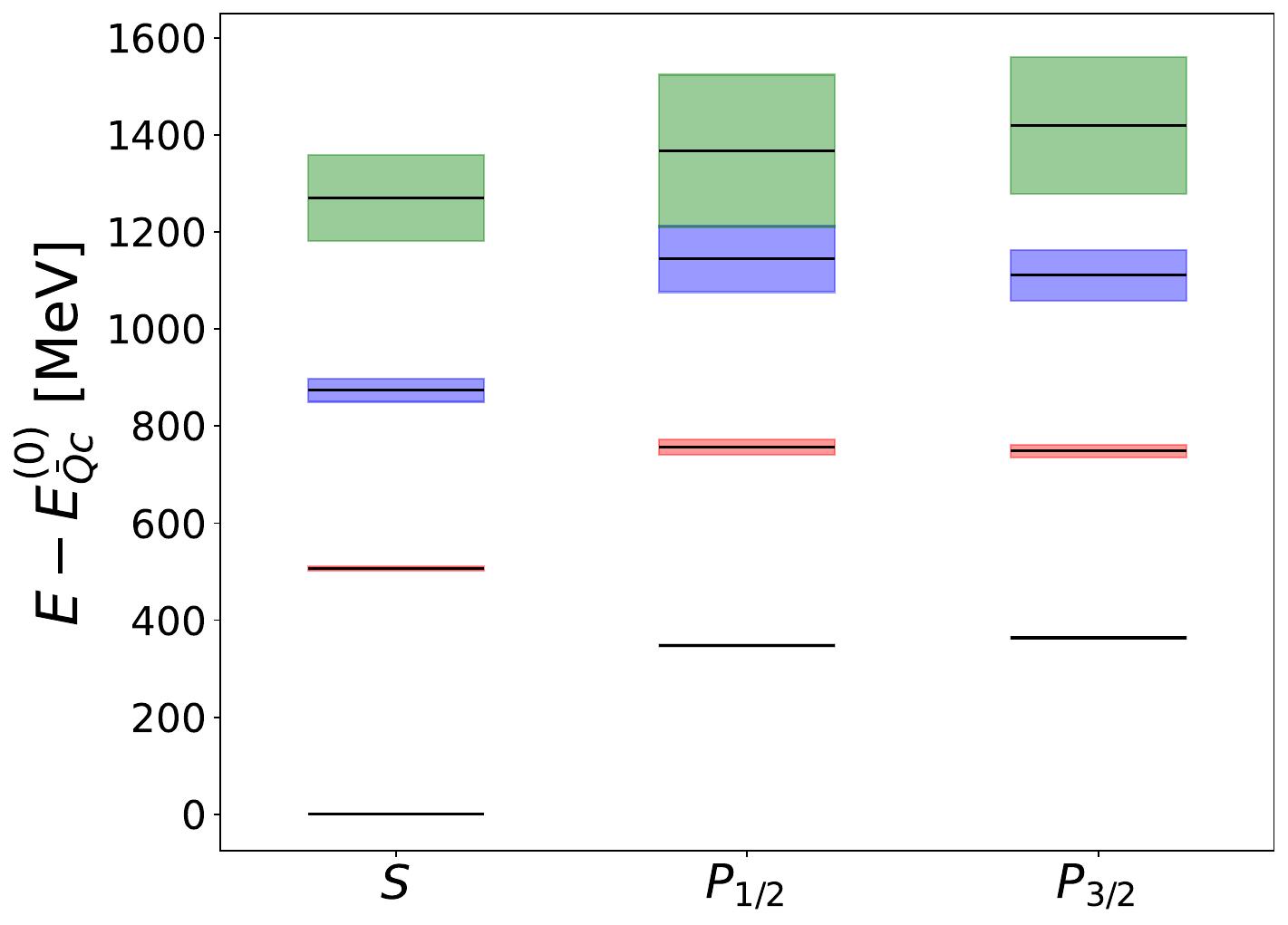}
    \caption{A1h ensemble.}
    \label{fig:second}
\end{subfigure}
\caption{Static-charm meson spectrum with accessible radial and orbital excitations, obtained using improved distillation.}
\label{fig:StatCharmSpectrum}
\end{figure}

Spectral measurements for the static-light and static-charm meson on the A1 and A1h ensemble in all given symmetry channels from table \ref{tab:SubductionTable} were carried out. The results are presented in figures \ref{fig:StatLightSpectrum} and \ref{fig:StatCharmSpectrum}. It is possible to construct $B^\ast \pi$ states with or without momentum of the pion to obtain the same quantum numbers as the static-light meson in the different symmetry channels. Thus, also non-interacting $B^\ast \pi$ energies $E_{B^{\ast}\pi}(|\Vec{p}|)$ with different pion momenta $\Vec{p}$ are included in the static-light spectrum in figure \ref{fig:StatLightSpectrum} for comparison. Due to the heavy quark spin symmetry, the mass of the vector static-light meson $B^\ast$ is degenerate with the mass of the pseudoscalar $B$, and thus both can be represented by $G_1^+$. To obtain the correct spin quantum numbers to compare with the static-light spectrum, the pion has to have momentum in certain cases. The subduction of pion representations for the smallest lattice momenta $|\Vec{p}|$ is given in table \ref{tab:PionRepr} \cite{Foley_2007}.
\begin{table}[H]
        \centering
        \begin{tabular}{c|c}
            $\Vec{p}$ & Irreducible content \\\hline
            $(0,0,0)$ & $A_{1}^-$ \\
            $(1,0,0)$ & $A_{1}^- \oplus E^- \oplus T_{1}^+$ \\
            $(1,1,0)$ & $A_{1}^- \oplus E^- \oplus T_{1}^+ \oplus T_{2}^+ \oplus T_{2}^-$
        \end{tabular}
        \caption{Subduction of pion representations with different units of momentum $|\Vec{p}|$ to the lattice irreps of $O_h^D$.}
        \label{tab:PionRepr}
\end{table}
According to this table, the $B^\ast$ has to be combined with a pion with one unit of momentum that transforms as $T_1^+$, since $G_1^+ \otimes T_1^+ = G_1^+ \oplus H^+$, to enable a projection onto the $G_1^+$ channel. For the $G_1^-$ ($P_{1/2}$) channel, the pion without momentum that transforms as $A_1^-$ can be taken, and for the $H^-$ ($P_{3/2}$), the pion with one unit of momentum that transforms as $E^-$ is needed. Given the correct momenta, the energy of the non-interacting $B^\ast\pi$ is $E_{\Bar{Q}l} + E_{\pi}$ with $E_{\Bar{Q}l}$ the energy of the static-light meson. For the pion energy $E_{\pi}$ the relativistic lattice dispersion relation
\begin{align*}
        \cosh{(aE_{\pi}(|\Vec{p}|))} = \cosh{(am_{\pi})}+\sum\limits_{k=1}^3\left(1-\cos{\left(a\frac{2\pi |n_k|}{L}\right)}\right)
\end{align*}
is used.


The magenta-colored bars in figure \ref{fig:StatLightSpectrum} correspond to these energies in all given channels. Although the pion mass is heavier in the A1h ensemble, the $B^\ast\pi$ states here lie closer to the measured level splittings than for the A1 ensemble, where they do not overlap with the measured splittings. Therefore, the A1 results appear to be more likely radial excitations of the $B$-meson and not $B^\ast\pi$ states. For a precise investigation of this excited state contamination, one needs to include $B^\ast\pi$ operators into the basis.
\\ The depicted static-light meson mass splittings show a significant dependence on the light-quark mass, whereas the level splittings of the static-charm meson, shown in figure \ref{fig:StatCharmSpectrum}, only exhibit a slight dependence, also shown in table \ref{tab:Splittings}. The reason for that is the subtraction of the ground state mass, which cancels out to a large extent the dependence on the charm quark mass in the case of the static-charm. For the $1P_{3/2}-1S$ splitting of the static-light meson and the $2S-1S$ splitting of the static-charm meson, there are experimental values given by the PDG \cite{PDG} that are also included in the table. Since the light mesons are degenerate in all considered ensembles, the $1P_{3/2}-1S$ splitting of the static-light meson can be compared to the corresponding splitting of the B-meson and the B$_s$-meson, as shown in the table. The values determined here do not agree with the PDG values, which are larger. However the correct trend to larger values when the pion mass is lowered towards the physical point can be observed. Even in the static limit an agreement with the PDG values is expected up to $1/m_b$ corrections when extrapolating the splittings to the physical pion mass. Finally, the masses determined here have a precision that is sufficient to calculate the splitting between the $1P_{3/2}$ and $1P_{1/2}$ states. The results are given in table \ref{tab:3/2-1/2Splitting}. Again, the splitting for the static-light shows a high dependence on the pion mass while it lies closer in the case of the static-charm.

\begin{table}
\renewcommand{\arraystretch}{1.1}
        \centering
        \begin{tabular}{c||c|c|c|c}
           & Splitting & $m_{\pi}\approx 420$ & $m_{\pi}\approx 800$ & PDG \\\hline\hline
           \multirow{3}{*}{static-light} & $1P_{1/2}-1S$  & $277.9(6.9)$ & $369.9(5.6)$ \\
           & $1P_{3/2}-1S$ & $408(10)$ & $384.7(5.3)$ & $m_{B_{2}^{\ast +}}-m_{B^0}  = 457.5(0.7)$ \\
           & & & & $m_{B_{s2}^\ast (5840)^0}-m_{B_s^0} = 473.0(0.2)$ \\
           \cline{2-5}
           & $2S-1S$ & $598.2(9.7)$ & $571(13)$ & \\\hline\hline
           \multirow{3}{*}{static-charm} & $1P_{1/2}-1S$  & $373.6(4.2)$ & $347.5(2.2)$ \\
           & $1P_{3/2}-1S$ & $401.1(4.5)$ & $363.7(2.4)$ & \\
           \cline{2-5}
           & $2S-1S$ & $521.8(6.2)$ & $506.3(4.3)$ &$m_{B_c (2S)^\pm}-m_{B_c^+} = 596.7(1.1)$ \\
        \end{tabular}
        \caption{Mass splittings between different excitations and the $1S$ ground state for the static-light and static-charm meson for both ensembles. All values are given in $\text{MeV}$.}
        \label{tab:Splittings}
\end{table}

\begin{table}
\renewcommand{\arraystretch}{1.1}
        \centering
        \begin{tabular}{c||c|c}
           $1P_{3/2}-1P_{1/2}$ splitting & $m_{\pi}\approx 420$ & $m_{\pi}\approx 800$ \\\hline\hline
           static-light & $130(12)$ & $14.8(7.6)$ \\\hline
           static-charm & $27.4(1.5)$ & $16.2(1.4)$ \\
        \end{tabular}
        \caption{Splitting between the $1P_{3/2}$ and the $1P_{1/2}$ for the static-light and static-charm meson for both ensembles. All values are given in $\text{MeV}$.}
        \label{tab:3/2-1/2Splitting}
\end{table}

\section{Conclusion and Outlook}
The improved distillation technique was applied and tested for static-light and static-charm mesons on two ensembles with different pion masses. It was shown that improved distillation enhances the mass plateaus compared to standard distillation. A larger number of eigenvectors is needed to give good access to excited states. By using improved distillation, the energy spectra for different radial and orbital excitations of the static-light and static-charm meson were presented, and a first naive approach was made to test the $B^\ast\pi$ excited state contamination in the static-light meson system. To discover possible $B^\ast\pi$ states in the static-light spectrum, interpolating operators that excite such states will be included in the operator basis in the next step. Since there is a degeneracy between the $H$ and the $G_2$ irreps that form the $5/2$ state, a measurement of $G_2$ will be performed to further identify radial excitations in the $H$ channel and investigate lattice artifacts. A study of the dependence of the mass splittings on the pion mass showed that the static-light meson spectrum is more sensitive to the light quark masses than the static-charm meson spectrum. Given the complete spectral analysis of one static-light meson, the next step towards hybrid string-breaking, considering two static-light mesons, will be performed.

\acknowledgments The authors gratefully acknowledge the Gauss Centre for Supercomputing e.V. (www.gauss-centre.eu) for funding this project by providing computing time on the GCS Supercomputer SuperMUC-NG at Leibniz Supercomputing Centre (www.lrz.de) under GCS/LS project ID pn29se as well as computing time and storage on the GCS Supercomputer JUWELS at J\"ulich Supercomputing Centre (JSC) under GCS/NIC project ID HWU35. The authors also gratefully acknowledge the scientific support and HPC resources provided by the Erlangen National High Performance Computing Center (NHR@FAU) of the Friedrich-Alexander-Universit\"at Erlangen-N\"urnberg (FAU) under the NHR project k103bf. M.P. was supported by the European Union’s Horizon 2020 research and innovation programme under grant agreement 824093 (STRONG-2020). R.H. was supported by the programme "Netzwerke 2021", an initiative of the Ministry of Culture and Science of the State of Northrhine Westphalia, in the NRW-FAIR network, funding code NW21-024-A. J. A. Urrea-Niño also acknowledges financial support by the Inno4scale project, which received funding from the European High-Performance Computing Joint Undertaking (JU) under Grant Agreement No. 101118139. The work is supported by the German Research Foundation (DFG) research unit FOR5269 "Future methods for studying confined gluons in QCD".

\bibliographystyle{JHEP}
\bibliography{ref.bib}

\end{document}